\documentclass[aps,twocolumn,pra,superscript,floatfix,superscriptaddress,showpacs,footinbib,reprint]{revtex4-1}
\usepackage{amssymb}

\usepackage[pdftex]{graphicx}
\usepackage{dcolumn}% Align table columns on decimal point
\usepackage{bm}% bold math
\usepackage{amsmath}
\usepackage{nicefrac}
\usepackage{array}
\usepackage{color}
\usepackage{float}
\usepackage{subfigure}
\usepackage{dsfont}
\usepackage{txfonts}
\usepackage{wasysym}
\usepackage{multirow}
\usepackage{sidecap}
\usepackage{xcolor,cancel}
\usepackage[plainpages=false,pdfpagelabels,colorlinks=true,linkcolor=red,urlcolor=blue,
citecolor=blue,pdftitle={arxiv-soc1},pdfauthor={},pdfdisplaydoctitle=true]{hyperref}
\usepackage[normalem]{ulem}

\newcommand{\+}{\dagger}

\newcommand{\dn}{\downarrow}

\newcommand{\e}{\varepsilon}

\newcommand{\wire}{\mathrm{wire}}

\newcommand{\hyb}{\mathrm{hyb}}

\newcommand{\up}{\uparrow}

\newcommand{\vecB}{\mathbf{B}}

\newcommand{\imp}{{\rm imp}}

\renewcommand{\Im}{{\rm Im \,}}

\newcommand{\HC}{{ \rm H.c.}}

\newcommand{\Bx}{\vecB \parallel \hat{\bf x}}
\newcommand{\By}{\vecB \parallel \hat{\bf y}}

\begin{document}

\title{Anisotropic Kondo screening induced by spin-orbit coupling in quantum wires}

\author{E. Vernek}
\affiliation{Instituto de F\'isica, Universidade Federal de Uberl\^andia, 
Uberl\^andia, Minas Gerais 38400-902, Brazil.}

\affiliation{Department of Physics and Astronomy, and Nanoscale and Quantum 
Phenomena Institute, Ohio University, Athens, Ohio 45701-2979, USA}

\author{G. B. Martins}
\affiliation{Instituto de F\'isica, Universidade Federal de Uberl\^andia, 
Uberl\^andia, Minas Gerais 38400-902, Brazil.}
\email[Corresponding author: ]{gbmartins@ufu.br}

\author{R. \v Zitko}
\affiliation{Jo\v zef Stefan Institute, Jamova 39, SI-1000 Ljubljana, Slovenia} 
\affiliation{Faculty of Mathematics and Physics, University of Ljubljana, Jadranska 19, SI-1000 Ljubljana, Slovenia.}

\date{\today}
\begin{abstract}
	Using the numerical renormalization group (NRG) method we study a magnetic impurity coupled 
	to a quantum wire with Rashba and Dresselhaus spin-orbit
	coupling (SOC) in an external magnetic field. We consider the low-filling regime 
  with the Fermi energy close to the bottom of the band and
  report the results for local static and dynamic properties in the Kondo regime.
	In the absence of the field, local impurity properties remain isotropic in spin space 
  despite the SOC-induced magnetic anisotropy of the conduction band.
  In the presence of the field, clear fingerprints of anisotropy are revealed
	through the strong field-direction dependence of the impurity
	spin polarization and spectra, in particular of the Kondo peak height.
  The detailed behavior depends on the relative magnitudes of the impurity and 
  band $g$-factors.
  For the case of impurity $g$-factor somewhat lower than the band $g$-factor,
  the maximal Kondo peak supression is found for field oriented along
	the effective SOC field axis, while for a field perpendicular to this
	direction we observe a compensation effect (``revival of the Kondo peak''):
	the SOC counteracts the Kondo peak splitting effects of the
	local Zeeman field. 
	We demonstrate that the SOC-induced anisotropy, measurable by
	tunneling spectroscopy techniques, can help to determine the ratio of 
	Rashba and Dresselhaus SOC strengths in the wire. 
\end{abstract}

\maketitle

\section{Introduction} 
The emergence of spin-orbit coupling as a major design principle in the development 
of new information technologies~\cite{DasSarma2004,
Bader2010}, especially after the discovery of topological insulators~\cite{Hasan2010}, 
has intensified studies of systems where SOC is determinant in 
providing access to the spin degree of freedom \cite{Winkler2003,Manchon2015}.
One of the main objectives is to incorporate spintronic ideas into  
contemporary technologies, which are overwhelmingly reliant on semiconducting materials~\cite{Fabian2007}. 
In this new paradigm, one aims spin injection, manipulation and detection using 
semiconductor structures similar to those already in widespread use in
standard semiconductor electronics.
Electron correlations and SOC may combine to produce 
new emergent behavior~\cite{Pesin2010,Balents2014,Rau2016,Schaffer2016}, as e.g. in iridates, $\rm{Sr_2IrO_4}$~\cite{Kim2008}.
The sensitivity of the Kondo effect~\cite{Bulla2008,Hewson}, the quintessential many-body phenomenon,
to magnetic anisotropy \cite{Romeike2006,Romeike2006b,Roosen2008,Otte2008,zitko2008,zitko2009,Pletyukhov2010,zitko2010,Misiorny2011,Misiorny2011b,zitko2011sc,Hock2013,Blesio2019} provides opportunities for novel devices.
In this work, the authors use the numerical renormalization group (NRG) method~\cite{Bulla2008} to 
study in unbiased manner an impurity in the Kondo regime under the combined 
effect of SOC~\cite{Meir1994,Malecki2007,Zitko2011,Zarea2012,Mastrogiuseppe2014,Wong2016,Chen2016,Sousa2016,Chen2017} and external magnetic field. 
More specifically, we consider a magnetic impurity in contact with a one-dimensional (1D) quantum wire, which is
subjected to Rashba~\cite{Bychkov1984} and
Dresselhaus~\cite{Dresselhaus1955} SOC, with the Fermi energy placed close to the bottom of the band, which is the regime
relevant for some of the proposed applications \cite{Kitaev2001,Oreg2010,Lutchyn2010,Nadj-Perge2014,Lutchyn2018,es2019}.

\begin{figure}[h]
	\centering
	\subfigure{
		\includegraphics[clip,width=0.8\columnwidth]{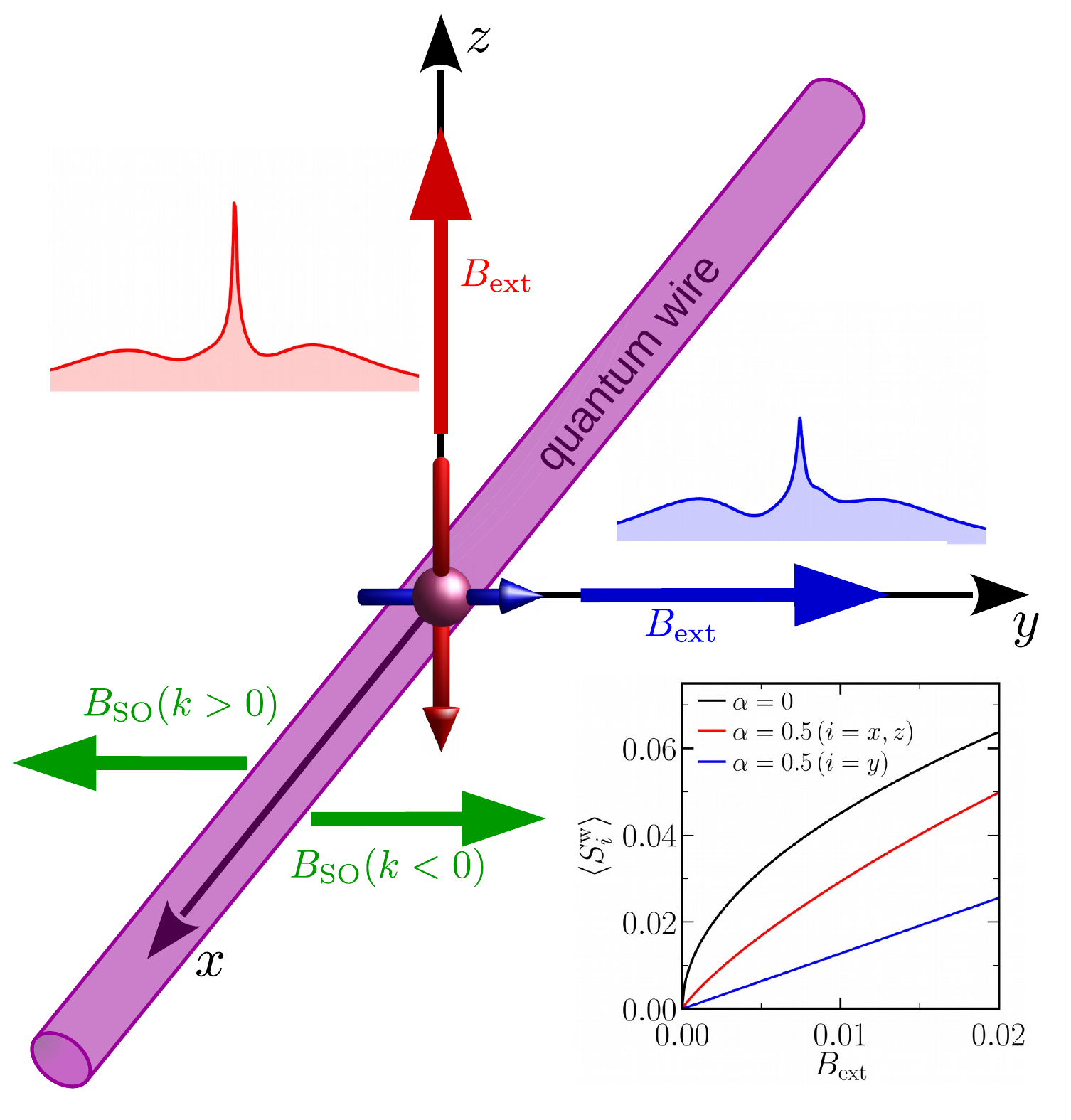}
	}
	\caption{\label{figcartoon} Sketch of the main results. 
	The impurity (purple sphere) is coupled to the quantum wire
	(purple line). The antiparallel effective $\vecB_{\rm SO}$
	magnetic field (green arrows) acts along the $y$-direction.
	The situation depicted is that for $g_{\rm imp} = 0.4 g_{\rm w}$, 
	with the impurity more weakly coupled to the external magnetic 
	field $\vecB$ than the quantum wire. 
	When $\vecB$ is applied along the $y$-axis (large blue arrow), 
	the Kondo peak is suppressed and split (blue
	curve),with the impurity polarized parallel to $\vecB$ (small blue arrow). 
	When $\vecB$ is applied along the $z$-axis (red
	arrow), the Kondo peak is more robust (red curve), yet the
	impurity polarization is stronger (small red arrow) and  
	furthermore it points in a direction opposite to $\vecB$. 
	Inset: spin
	polarization of the wire, $\langle S_i^{\rm w} \rangle$, as a function of 
	$\vecB$ applied along the 
        different axes. The black curve corresponds to zero SOC, $\alpha=0$. 
	} 
\end{figure}

The main result is sketched in Fig.~\ref{figcartoon}. 
The wire is oriented along the $x$-axis and, for simplicity, pure
Rashba SOC is considered here, hence the effective SOC magnetic field
$\vecB_{\rm SO}$ (antiparallel green arrows) points along the
$y$-axis. An external magnetic field acts on both the impurity and the wire with different $g$ factors,
denoted as $g_\mathrm{imp}$ and $g_w$, respectively. 
In order to probe the physical origins of the various contributions to the impurity 
total spin polarization, we consider two cases, viz., one with $g_\mathrm{imp}=0$ 
and another with $g_\mathrm{imp} \neq 0$.
We consider the case of $T \ll T_K$, where $T_K$ is the Kondo temperature for finite-SOC and vanishing external magnetic field $\vecB$. 
If $\vecB$ points along the $y$-axis (red arrow), the Kondo peak is suppressed 
(red sketch). 
However, for $\vecB$ along the $x$ or $z$-axis (blue arrow), the
Kondo peak persists (blue sketch). 
The {\it impurity} spin polarization is also anisotropic: for $\vecB$
along $y$-axis, the impurity is only slightly polarized in the direction of $\vecB$ 
(horizontal red arrow), while for $\vecB$ along $x$ or $z$
axis the impurity is considerably more polarized, but the
spin polarization is oriented opposite to $\vecB$ (vertical blue arrow). 
One might be led to expect that the stronger suppression of the Kondo peak 
for $\vecB$ applied along the $y$-axis 
implies stronger polarization of the quantum wire when the external 
magnetic field is applied along this direction. 
However, this is not the case: the inset to Fig.~\ref{figcartoon} shows that
in the presence of SOC  the {\it wire} spin polarization, $\langle S_i^{\rm w}
\rangle$~\cite{note7}, is always reduced compared
to the zero-SOC case, but the supression is actually greater for the
case of external field along the effective SOC-field direction.
The stronger suppression of the Kondo peak for $\vecB$ along
the $y$-axis hence cannot be explained by the polarization of the conduction electrons and one
instead needs to consider dynamic effects, as we do in the following.

\section{Model and Hybridization Function}\label{sec-model} 
\subsection{Model}
The wire Hamiltonian is 
\begin{eqnarray}\label{H_wire}
H_\wire&=\sum_k \Psi^\dagger_k{\cal H}_\wire \Psi_k, \\
{\cal H}_\wire&=\left(\e_k -\mu\right)\sigma_0 + \vecB_{\rm tot}\!\cdot{\bm \sigma}.
\end{eqnarray}
Here $\Psi^{\dagger}_k=(c^\dagger_{k\up}, c^\dagger_{k\dn})$, $c^\dagger_{k\sigma}$ creates 
an electron with wave vector $k$ and spin $\sigma=\uparrow,\downarrow$,
$\e_k=-2t\cos k$ is the tight-binding dispersion relation where
$t$ is the nearest-neighbor hopping matrix element, $\mu$ is the
chemical potential, $\vecB_{\rm tot}\!=g_{\rm w} \vecB+\vecB_{\rm SO}$\!
represents the combined effect of an external magnetic field $\vecB=(B_x,B_y,B_z)$ 
and an effective $k$-dependent spin-orbit magnetic field~\cite{Manchon2015} 
${\bf B}_{\rm SO}(k)=\sin k(\beta,-\alpha,0)$, where the couplings $\alpha$ and $\beta$ 
(measured in energy units) are the Rashba~\cite{Bychkov1984} and Dresselhaus~\cite{Dresselhaus1955} SOC strengths, 
respectively.
The vector of Pauli matrices ${\bm \sigma} = \{\sigma_x,
\sigma_y,\sigma_z\}$ and the identity matrix 
$\sigma_0$ act on spin space. For simplicity, we set the Bohr magneton
to $\mu_B = 1$, and the factor $\nicefrac{1}{2}$ from ${\bm S}=\nicefrac{\bm \sigma}{2}$
has been absorbed into $g_{\rm w}$. 
We parameterize both SOCs as
$\theta_{\rm SO} = -\tan^{-1}\nicefrac{\alpha}{\beta}$, such that 
$\beta=\gamma \cos \theta_{\rm SO}$ and $-\alpha=\gamma \sin
\theta_{\rm SO}$, i.e. $\theta_{\rm SO}$ is the angle between the effective magnetic field
${\bf B}_{\rm SO}(k)$ (for positive $k$) and the $x$-axis. 
For pure Rashba SOC with $\beta=0$ ($\theta_{\rm SO}=\pm\nicefrac{\pi}{2}$), 
the effective field points along the $y$-axis (see Fig.~\ref{figcartoon}), while
for pure Dresselhaus SOC with $\alpha=0$ ($\theta_{\rm SO}=0,\pi$), it
points along the $x$-axis. 

To study the Kondo state in this system, the quantum wire is coupled to an Anderson impurity, which is modeled as  
\begin{eqnarray}\label{eq:himp}
	H_\imp=\sum_s \e_d n_\sigma + Un_\up n_\dn + g_{\rm imp} \vecB\cdot{\bm S}, 
\end{eqnarray}
where $d^\+_{\sigma}$ ($d_{\sigma}$) creates (annihilates) an electron with orbital 
energy $\e_d$ and spin $\sigma=\uparrow, \downarrow$, $n_{\sigma}=d^\+_{\sigma}d_{\sigma}$, and 
$U$ represents Coulomb repulsion. The third term accounts for 
the Zeeman interaction of the impurity's magnetic moment $g_{\rm imp} {\bm S}$.
The hybridization between the impurity and the conduction electrons is given by
\begin{eqnarray}
	H_\hyb=\sum_{k\sigma}\left( V_k  d^+_{\sigma} c_{k\sigma}+\HC \right).
\end{eqnarray}
In this work we consider the case of $V_k \equiv V$.
The Fermi energy is close to the bottom of the band, 
$\mu=-1.0$, and we use the $\gamma=0$ half-bandwidth $D=2t=1.0$ as the energy unit. 
Unless stated otherwise, we use $U=0.5$, $\epsilon_d=-U/2$, $V=0.07$, and $g_{\rm w}/g_{\rm imp} = 2.5$~\cite{note3} 
(with $g_{\rm w} = 1$ and $g_{\rm imp}=0.4$), $|\vecB|=0.01$. We take $\gamma=\sqrt{\alpha^2 + \beta^2}=0.5$ as being fixed and
perform most calculations for $\theta_{\rm SO}=\nicefrac{-\pi}{2}$ (i.e., Rashba-only). 
The Kondo temperature of this system is $T_K \approx 1.16 \times 10^{-2}$\!.

\begin{figure}[h!]
\centering
\subfigure{\includegraphics[clip,width=1.0\columnwidth]{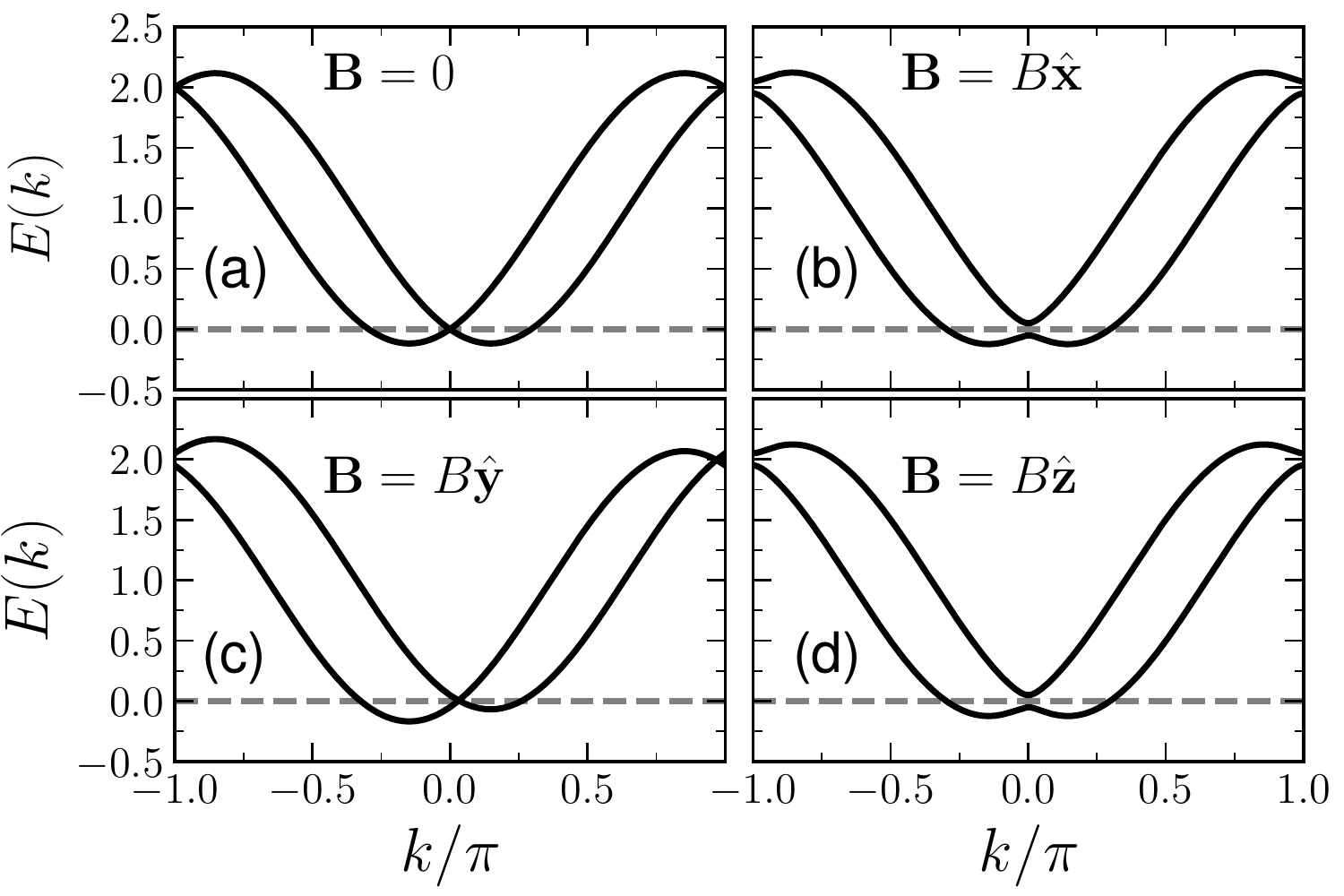}}
\caption{\label{fig1} Band structure of a quantum wire for
$\gamma=0.5$, $\theta_{\rm SO}=\nicefrac{-\pi}{2}$ (Rashba SOC)
	and different values of external field 
	(a) $B=0$, (b) $\vecB=B\hat{\mathbf{x}}$, (c)
	$\vecB=B\hat{\mathbf{y}}$, (d)
	$\vecB=B\hat{\mathbf{z}}$, with $B=0.01$. 
	}
\end{figure}

The single-electron bands described by {\cal H} are shown in
Fig.~\ref{fig1}: panel a for zero $\vecB$, and 
panels b,c,d for $|\vecB|=0.01$ oriented along $x$, $y$,
and $z$-axis. For $\theta_{\rm SO}=\nicefrac{-\pi}{2}$, ${\bf B}_{\rm SO}(k)$ 
is oriented along the $y$-axis, thus the bands for $\vecB$ along the $x$ and $z$-axis
are identical, but they differ from those for $\vecB$ along
the $y$-axis, reflecting the anisotropy introduced by SOC. 
The dependence of the direction of $\vecB_{\rm SO}$ on the sign of $k$
can be read from the difference in the splitting of the bands in panel c: 
for $k>0$ $\vecB_{\rm SO}$ opposes $\vecB$, generating a smaller 
band splitting, while for $k<0$ $\vecB_{\rm SO}$ aligns with $\vecB$, 
increasing the band splitting. In the other cases the bands have even
parity.

\subsection{Hybridization function} 
The impurity Green's function $\hat G_\imp(\omega)$ can be written as
\begin{eqnarray}\label{Gimp}
	\hat G_\imp(\omega)=\left[\left(\omega -\e_d\right)\sigma_0 -
	\hat{\Sigma}^{\rm (int)}(\omega) -
	\hat{\Sigma}^{(0)}(\omega)\right]^{-1},
\end{eqnarray}
where $\hat{\Sigma}^{\rm (int)}(\omega)$ is the interaction self-energy,
while $\hat{\Sigma}^{(0)}(\omega)=\sum_k \hat V \hat G_\wire(k,\omega) \hat V^\+ $ is the hybridisation self-energy,
with $\hat V=V_0\sigma_0$ and $\hat G_\wire(k,\omega)=\left[\omega \sigma_0-{\cal H}_\wire \right]^{-1}$.
One finds
\begin{equation*}
\begin{split}
\hat{\Sigma}^{(0)}(\omega) = \sum_k F(k,\omega)\Bigl[
\left(\cos k+\mu +\omega\right)\sigma_0 + \\
(\alpha  \sigma_y-\beta  \sigma_x)\sin k -g_{\rm w}\vecB\cdot {\bm \sigma}  \Bigr],
\end{split}
\end{equation*}
where
\begin{eqnarray*}
F(k,\omega)=\frac{-V^2}{2  \left(\alpha  B_y -\beta B_x \right)\sin k+B^2+\gamma^ 2 \sin ^2k-(\cos k+\mu +\omega )^2}.
\end{eqnarray*}
For a magnetic field applied along an arbitrary direction, 
$\hat \Sigma^{(0)}(k,\omega)$ has finite off-diagonal terms and
we have to deal with a spin-mixing hybridization function~\cite{Liu2016,Osolin2017}
\begin{eqnarray}
\hat{\Gamma}(\omega)=\frac{1}{2i}\int_{-\pi}^{\pi}\left[\hat\Sigma^{(0)}(k,\omega-i0^-) - \hat\Sigma^{(0)}(k,\omega+i0^+) \right] dk. 
\end{eqnarray}
This positive-definite Hermitian matrix can be decomposed
in terms of Pauli matrices as $\hat\Gamma(\omega)=\sum_{i\in \{0,x,y,z\}} d_i(\omega)\sigma_i$, 
where all $d_i(\omega)$ are real quantities. In particular, $d_0(\omega)$ is proportional to the
conduction-band density of states~\cite{note0}. 
\begin{figure}[h]
\centering
\subfigure{
\includegraphics[clip,width=1.0\columnwidth]{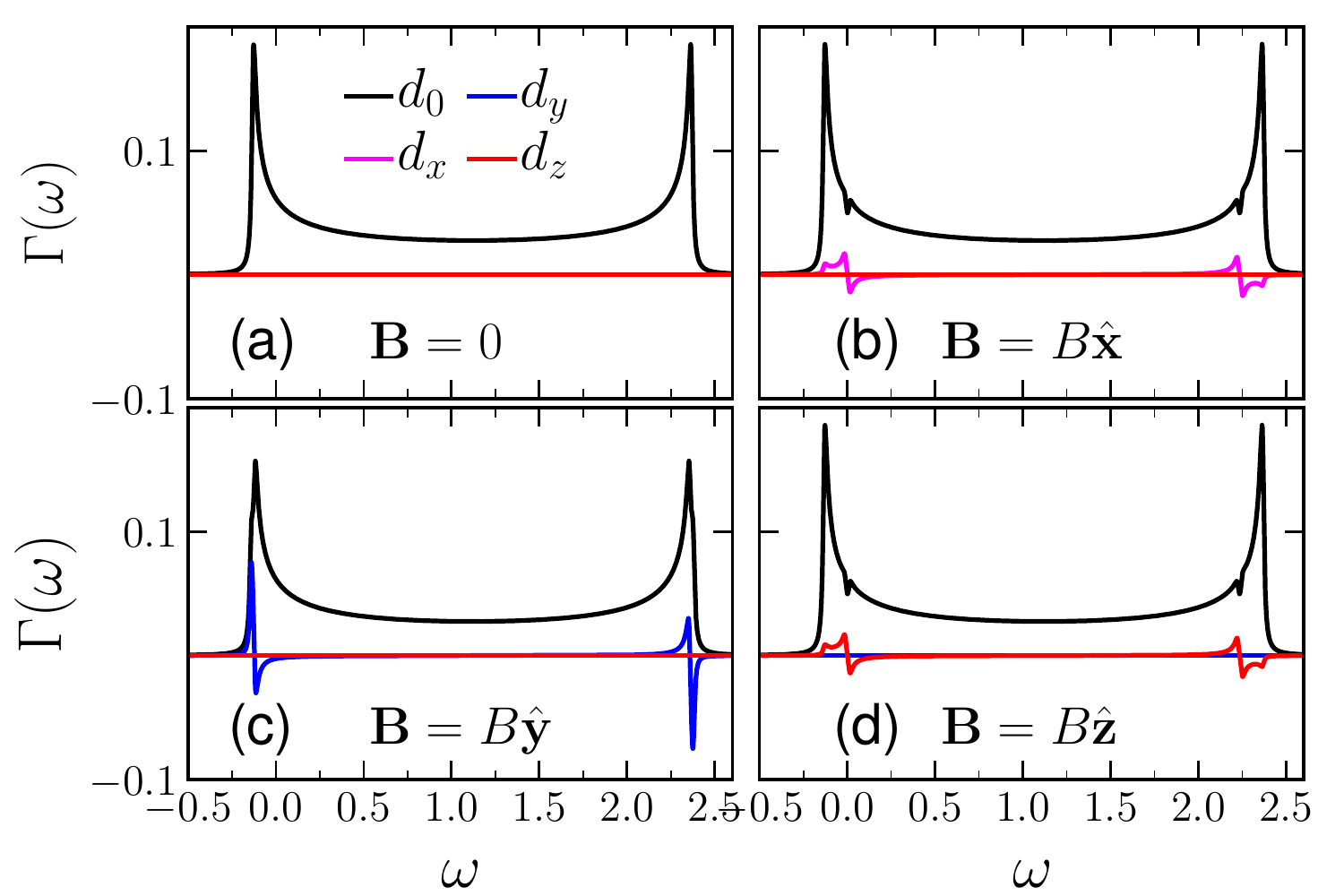}
}
\caption{\label{fig2} Hybridization function coefficients
$d_i(\omega)$ for (a) zero field, and for (b,c,d) field oriented along
the different axes.}
\end{figure} 
In the absence of SOC, 
for $B=0$, only $d_0(\omega)$ is non-zero, while for $B>0$, the
coefficient $d_i$ in the field direction is also finite, with a
value that does not depend on the field direction, thus manifesting
the spin isotropy.
In the presence of SOC, the rotation invariance is broken, see
Fig.~\ref{fig2}.
For $B=0$ (panel a) again only $d_0(\omega)$ is non-zero.
For $\vecB=B\hat{\mathbf{x}}$ (panel b), $d_0(\omega)$ exhibits a small dip 
associated to the lifting of degeneracies at $k=0$ and $\pi$ [see Fig.~\ref{fig1}(b)],
$d_x(\omega)$ is finite, while $d_y(\omega)$ and $d_z(\omega)$ remain
zero. The results in panel d, for the field along $z$-axis, are
equivalent up to a permutation of the $x$ and $z$ axes.
For $\vecB=B\hat{\mathbf{y}}$ (panel c), $d_0(\omega)$ is different from the corresponding curve in 
panels b and d, and $d_y(\omega)$ is different from $d_x(\omega)$ and $d_z(\omega)$ in those panels. 
This clearly shows  how the $y$-axis becomes distinct, since the Rashba SOC tends to align the spins of the conduction 
electrons along this axis.
The anisotropy of $\hat{\Sigma}^{(0)}$ affects the screening of
impurity local moment~\cite{Bulla2008}, thus the SOC in the wire is experimentally
detectable by probing the properties of the Kondo state. The problem
bears some similarity with the problem of a quantum dot with
ferromagnetic leads
\cite{Sindel2007,Choi2004,Martinek2005,Martinek2003a,Martinek2003b,Zitko2012},
but the focus here is on SOC anisotropy and the ensuing detailed form
of the hybridisation function, with complex (and direction-dependent) behavior 
close to the band edges.

\begin{figure}[h]
	\centering
	\subfigure{
		\includegraphics[clip,width=1.0\columnwidth]{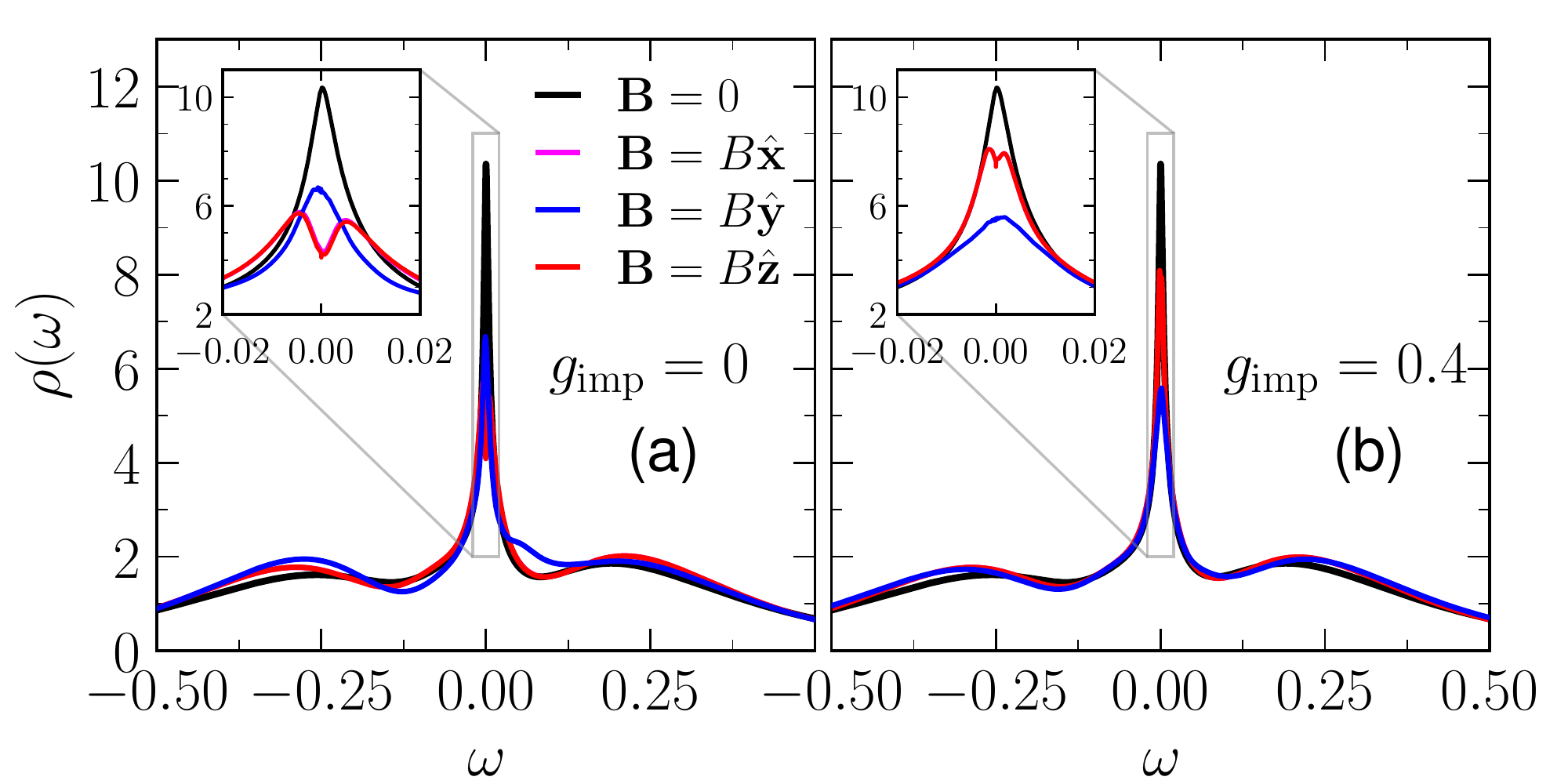}
	}
	\caption{\label{fig4} Local density of states $\rho(\omega)$.
	 (a) $g_{\rm imp}=0$ and (b) $g_{\rm imp}=0.4$. 
	Insets: close-ups on the Kondo peak. 
	}
\end{figure}

\section{Results}
\subsection{Impurity Local Density of States} 
The total impurity LDOS $\rho(\omega)=-\frac{1}{\pi} \Im \mathrm{Tr}\, \hat{G}_{\imp}(\omega)$
is shown in Fig.~\ref{fig4} for zero field and for fields along the three axis. 
Two different g-factor values are used: $g_{\rm imp}=0$ (panel a) and $g_{\rm imp}=0.4$ (panel b). 
The Kondo peak is similarly suppressed for $\By$ for both 
g-factor values, slightly more so for finite $g_{\rm imp}$ (see insets), while for $\Bx$ or $\hat{\bf z}$ there 
is a noticeable quantitative difference: the splitting and suppression 
is much more prominent for vanishing $g_{\rm imp}$.
Thus, the picture that emerges is the following: for $g_{\rm imp}=0$ the band polarization 
results in the inset to Fig.~\ref{figcartoon} explain the Kondo suppression for any direction 
of the external magnetic field. However, when the impurity Zeeman effect is turned on, the Kondo suppression 
for $\By$ is largely unaffected, while for $\Bx$ or $\hat{\bf z}$ it is partially \emph{erased}, 
as if the band polarization and the impurity Zeeman effect were canceling each other.
In other words, for $\Bx$ or $\hat{\bf z}$, a finite Zeeman term at the impurity ($g_{\rm imp}=0.4$, panel b) 
seems to partially compensate the broad splitting 
caused by the band polarization (panel a), as it increases the LDOS spectral weight around $\omega=0$, 
partially reconstructing the Kondo peak. 

\begin{figure}[h]
	\centering
	\subfigure{
		\includegraphics[clip,width=1.0\columnwidth]{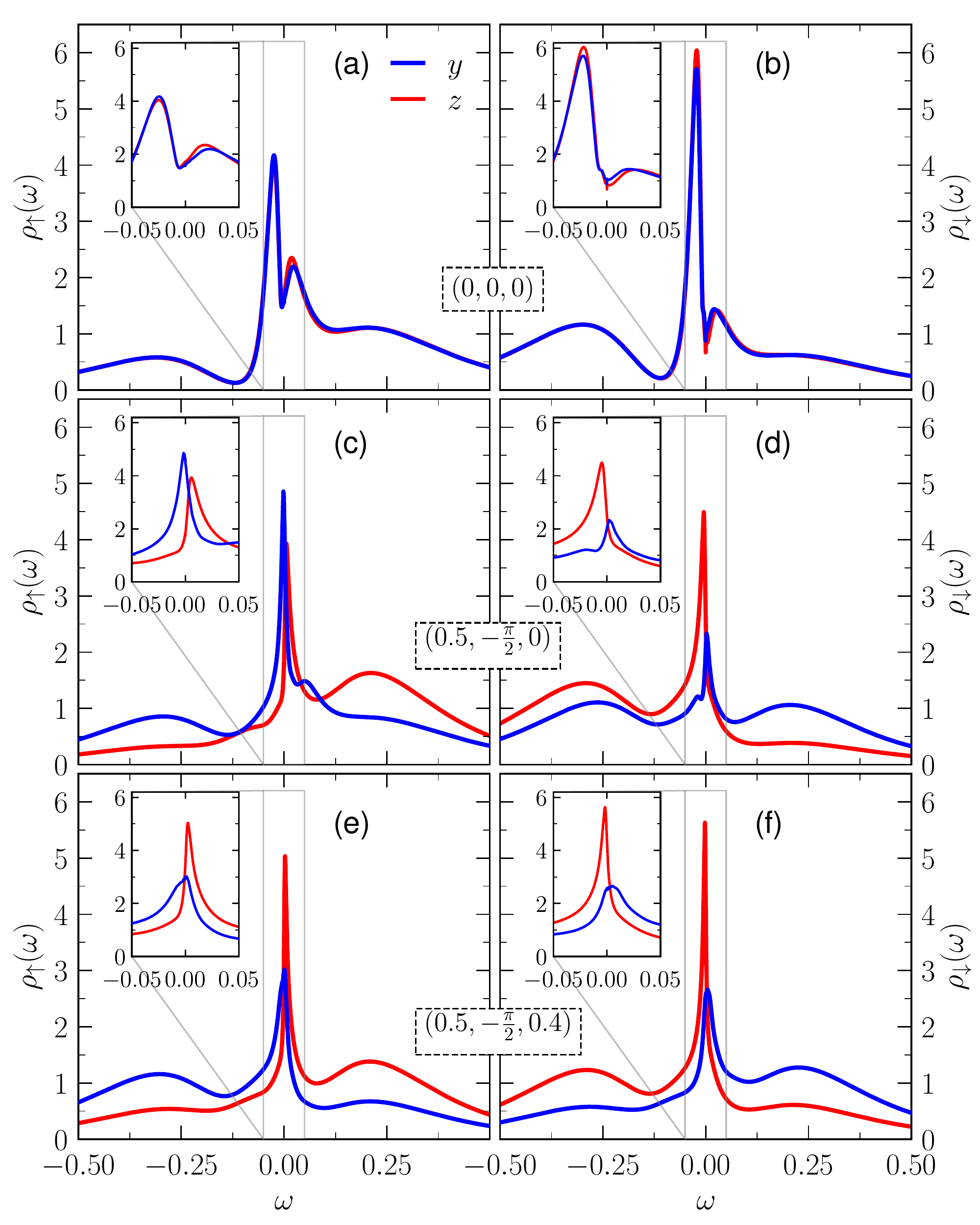}
	}
	\caption{\label{fig6} Spectral function resolved along the
	axis of the applied magnetic field, $\rho_{\sigma_i}(\omega)$,
	with $i \in \{x,y,z\}$;
	left and right panels correspond to the two projections.
	Top-row: zero SOC and $g_{\rm imp}=0$ (reference results).
        Middle-row: Rashba SOC [$\gamma=0.5$, $\theta_{\rm
	SO}=\nicefrac{-\pi}{2}$] and $g_{\rm imp}=0$. 
	Bottom-row: Rashba SOC and $g_{\rm imp}=0.4$. In all panels
	$g_{\rm w}=1.0$, i.e., the quantum wire is spin polarized. The insets show 
	a close-up of the vicinity of the Fermi energy. The tuples straddling both panels 
	for each row indicate the respective values of $[\gamma,\theta_{\rm SO},g_{\rm imp}]$.
		} 
\end{figure} 

\subsection{Spin-resolved Local Density of States.} 
\begin{figure}[h]
	\centering
	\subfigure{
		\includegraphics[clip,width=1.0\columnwidth]{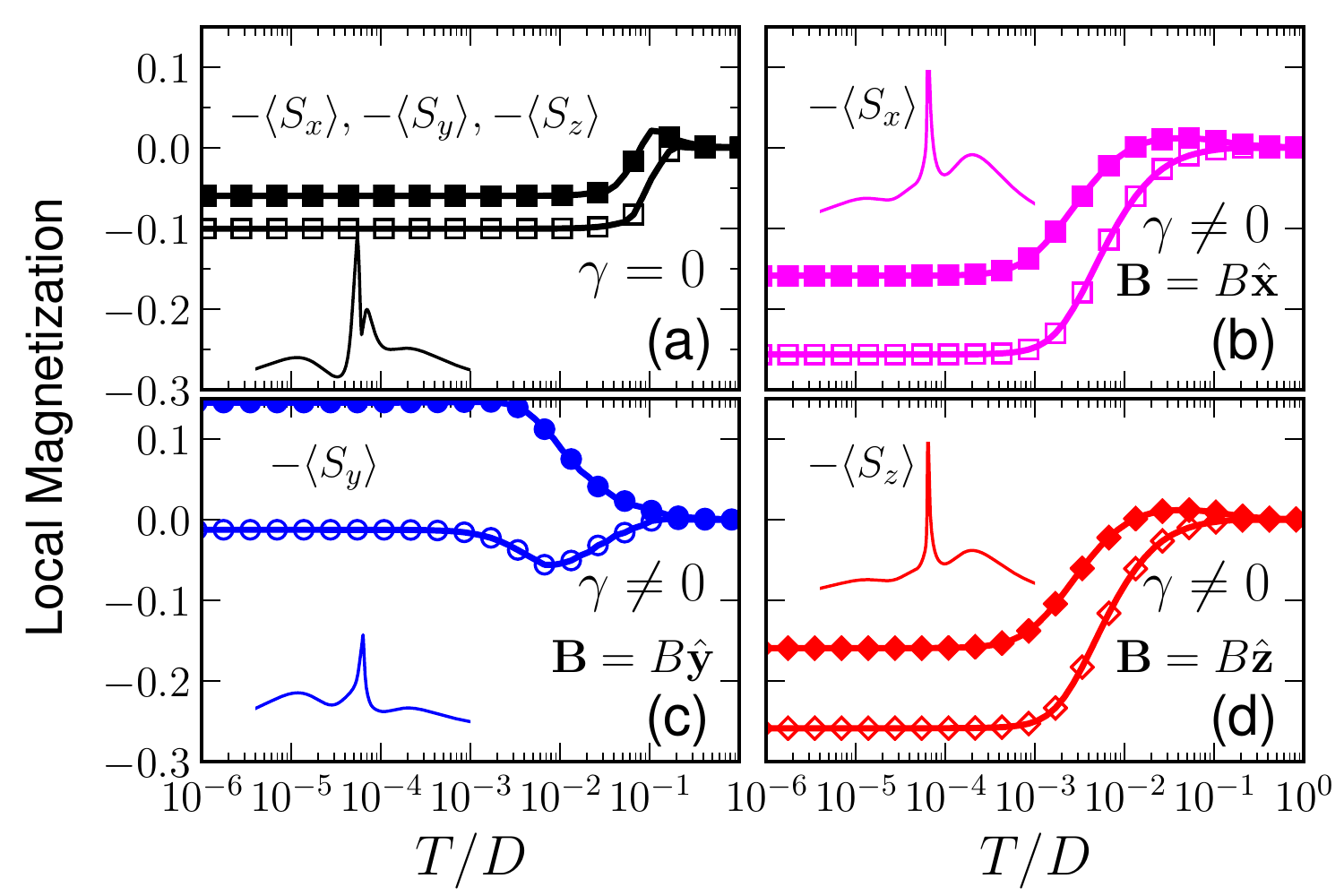}
	}
	\caption{\label{fig5} Impurity spin polarization $\langle S_i \rangle$ \emph{vs} 
	temperature. Same parameters as in Figs.~\ref{fig2} and \ref{fig4}, except 
	for panel a, where $\gamma=0$. In panels a to d, 
	open and solid symbols correspond to $g_{\rm imp}=0$ and
	$g_{\rm imp}=0.4$, respectively. Note that in panel a, since $\gamma=0$, $\langle S_i \rangle$ 
	is the same for all $i$ axes, thus only the $x$-axis result is shown. In panels b, c, and d, 
	$i=x$, $y$, and $z$, respectively. Note that the impurity
	polarization, for directions 
	perpendicular to the applied external field [like, for example, 
	$\langle S_y \rangle$ and $\langle S_z \rangle$, for $\Bx$, panel b], 
	vanish identically, and therefore are not shown. In each panel, a sketch of the impurity's LDOS,  
	corresponding to $g_{\rm imp}=0$, is shown. 
	} 
\end{figure}

We now reexamine the spectra 
by resolving them along the magnetization axis defined by the applied 
external magnetic field, see Fig.~\ref{fig6}. 
The three rows
of panels show the results as couplings are gradually turned on: 
(i) $\gamma=0$, $g_{\rm imp}=0$,
(ii) $\gamma=0.5$, $g_{\rm imp}=0$,
(iii) $\gamma=0.5$, $g_{\rm imp}=0.4$. For all rows $g_{\rm w}=1.0$. 
At zero SOC (first row), the
results do not depend on the field direction.
Since $g_{\rm imp}=0$, the suppression of the Kondo peak and the partial polarization of the impurity is 
induced by the band polarization alone.
In the presence of SOC (second row), $\By$ differs from $\Bx$ or $\hat{\bf z}$. 
In addition, since the introduction of SOC moves the van-Hove singularity at the bottom 
of the band to lower energies, away from the Fermi energy, the Kondo peak becomes less asymmetric
compared to the $\gamma=0$ results in the first row.
Resolving the spectra along the external field direction allows us to see that the Kondo 
effect is affected more strongly when $\By$, since, as is more clearly seen in the insets, the 
Kondo peak polarization is parallel to the applied field for $\By$ and antiparallel 
for $\Bx$ or $\hat{\bf z}$. The inclusion of a finite $g_{\rm imp}$ (third row) 
changes this picture only quantitatively, with the impurity 
becoming less antiferromagnetically correlated with the polarized band for $\Bx$ 
and $\hat{\bf z}$, and becoming more ferromagnetically correlated with the band 
for $\By$. This picture is reinforced by calculating the impurity polarization as 
a function of temperature, see Fig.~\ref{fig5}, where one can see 
that, at low temperature, for $\By$ and $g_{\rm imp}=0$ (open symbols in panel c), 
the impurity is barely correlated with the band, becoming ferromagneticaly correlated 
with it for $g_{\rm imp}=0.4$ (solid symbols). On the other hand, for $\Bx$ and $\hat{\bf z}$,  
(panels b and d), there is a clear Kondo correlation of the impurity with the band 
for $g_{\rm imp}=0$ (open symbols), which is somewhat weakened by the impurity Zeeman term 
($g_{\rm imp}=0.4$, solid symbols). Jointly, these results establish the revival alluded 
to in Fig.~\ref{figcartoon}, as moving the external field from $\hat{\bf y}$ to $\hat{\bf x}/\hat{\bf z}$ 
strengthens the Kondo effect (see Fig~\ref{fig7} too). 

\subsection{Temperature dependence of the impurity spin polarization} 
Now, we track the impurity spin polarization, $\langle S_i \rangle$~\cite{note9}, 
as the temperature is reduced from $T=D$ to $\approx 0$, see
Fig.~\ref{fig5}. By following
how the spin components evolve through the three SIAM fixed points, we
gain some intuition on how the SOC affects the Kondo state properties. In
addition, by comparing the results for $g_{\rm imp}=0$ (open symbols) and $g_{\rm
imp}=0.4$ (solid symbols), we discern which effects arise from the band polarization
alone, and which are the consequence of the local Zeeman field.
An external magnetic field $|\vecB|=0.01$ 
is applied along the same $i$-axis along which the impurity spin magnetization -$\langle S_i \rangle$ is measured. 
The results in panel a, without SOC ($\gamma=0$), are the same for all three directions
(thus, only the $x$-axis result is shown). 
The temperature variation of the impurity magnetization reveals the cross-overs between the three 
SIAM fixed points: free orbital (FO) $\rightarrow$ local moment (LM) $\rightarrow$ strong coupling (SC).  
At the FO fixed point ($T \lesssim D$), the spin magnetization is negligible for both values of 
$g_{\rm imp}$ because of the strong charge fluctuations.
At the LM fixed point, the local spin starts to form for $T \lesssim
U=0.5$ and the open and solid symbols curves start to separate: for $g_{\rm imp}=0$ the impurity 
polarizes in response to the band polarization and its spin antialigns
with the band polarization due to antiferromagnetic Kondo exchange coupling (thus $-\langle S_i \rangle < 0$), 
while for $g_{\rm imp}=0.4$ the impurity Zeeman term will counteract this effect (thus $-\langle S_i \rangle \gtrsim 0$). 
As the temperature decreases further ($T \approx U/5 = 0.1$), the charge fluctuations die down and, 
for $g_{\rm imp}=0.4$, $-\langle S_i \rangle$ reaches a maximum at the LM fixed point and decreases 
toward the SC fixed point. Because the Zeeman effect is too small to suppress Kondo,  
the magnetization settles into an $-\langle S_i \rangle < 0$ plateau located above that for $g_{\rm imp}=0$. 

The results for finite SOC are shown in panels b to d.
For $\By$ (panel c), by comparison to the results just described for 
zero-SOC ($\gamma=0$), we see that the combination of SOC and $\vecB \parallel \vecB_{\rm SO}$ 
considerably weakens the Kondo state resulting from finite $\vecB$ and $\gamma=0$ (panel a), 
since the $-\langle S_i \rangle \approx 0$ plateau for $g_{\rm imp}=0$ indicates that 
the impurity is barely correlated to the band, and $-\langle S_i \rangle > 0$ for $g_{\rm imp}=0.4$. 
On the other hand, for $\Bx$ or $\hat{\bf z}$ (panels b and d), where $\vecB \perp \vecB_{\rm SOC}$, 
the situation is quite different, as it is clear that the Kondo state was strengthened in relation to both the zero-SOC 
case (panel a) and the finite SOC with $\By$ case (panel c), illustrating 
the Kondo `revival' shown in Fig.~\ref{fig7}.

\subsection{Field dependence of impurity magnetization and Kondo splitting} 
\begin{figure}[h]
	\centering
	\subfigure{
		\includegraphics[clip,width=1.0\columnwidth]{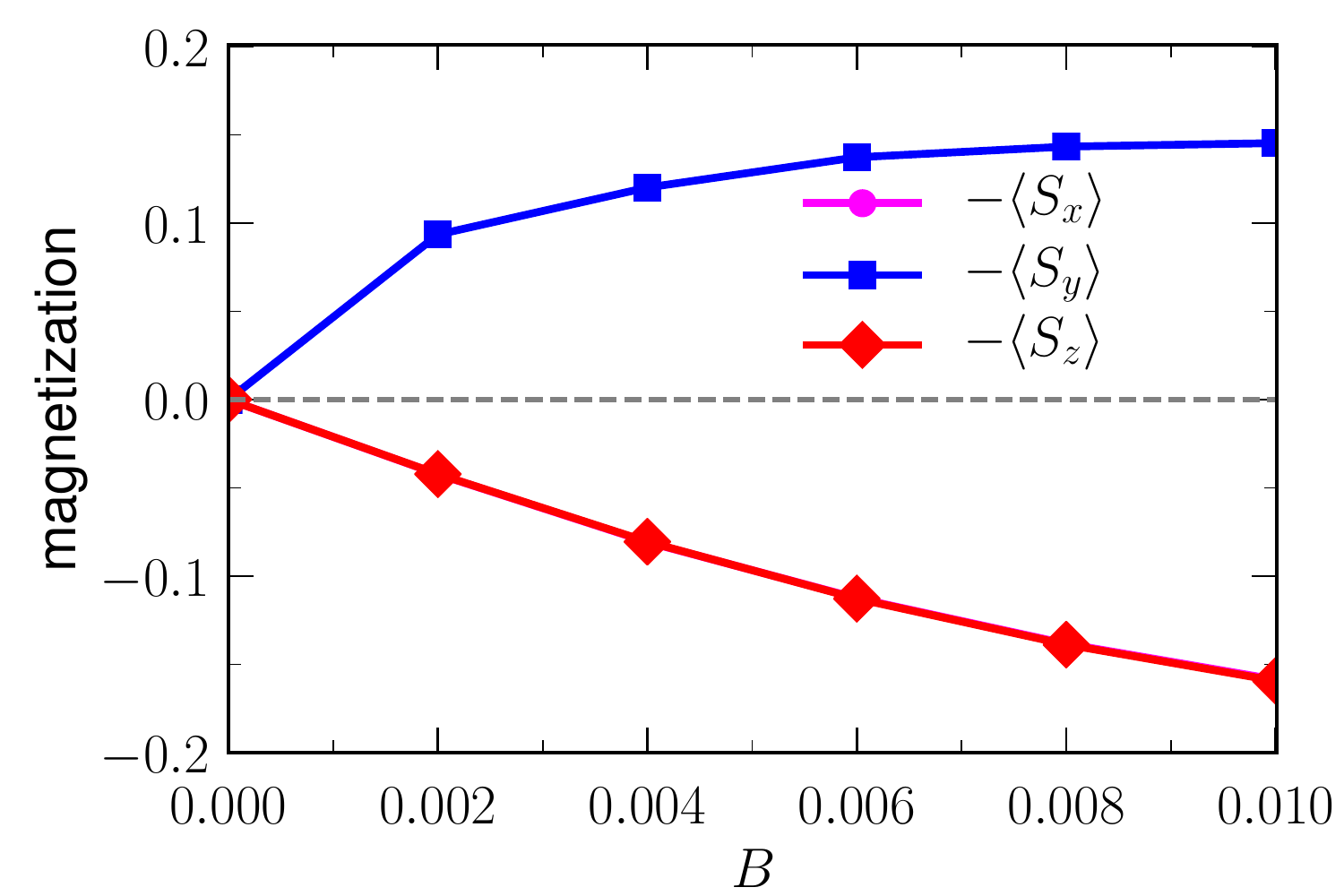}
	}
	\caption{\label{figmagvsB} Impurity spin magnetization $-\langle S_i
	\rangle$ \emph{vs} $|\vecB|$ for field along different directions, for $g_{\rm imp}=0.4$.
		} 
\end{figure}

\begin{figure}[h]
	\centering
	\subfigure{
		\includegraphics[clip,width=1.0\columnwidth]{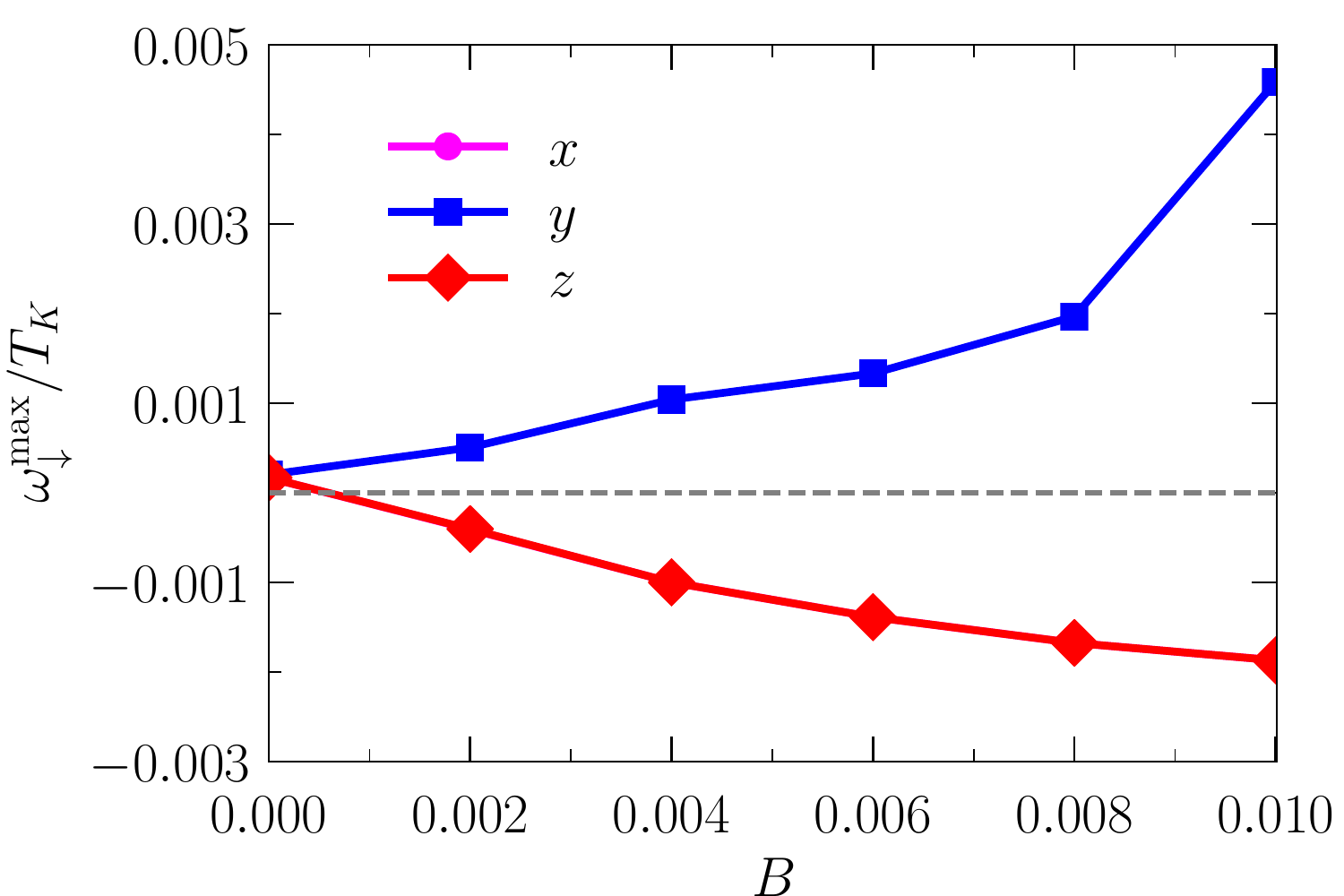}
	}
	\caption{\label{figKpeakvsB} Spin-down projected Kondo peak position 
	$\omega_{\downarrow}^{\rm max}$ as a function of $\vecB$, for $g_\mathrm{imp}=0.4$. 
	The impurity spin polarizes (magnetizes) along (opposite to) the external field for $\By$, while 
	the reverse occurs for $\Bx$ or $\hat{z}$. In other words, the impurity spin correlates 
	antiferromagnetically with the band spins in the latter case, and ferromagnetically for the former case. 
	This is in accordance with the result sketched in Fig.~\ref{figcartoon}.  
	} 
\end{figure}

\begin{figure}[h!]
	\centering
	\subfigure{\includegraphics[clip,width=1.0\columnwidth]{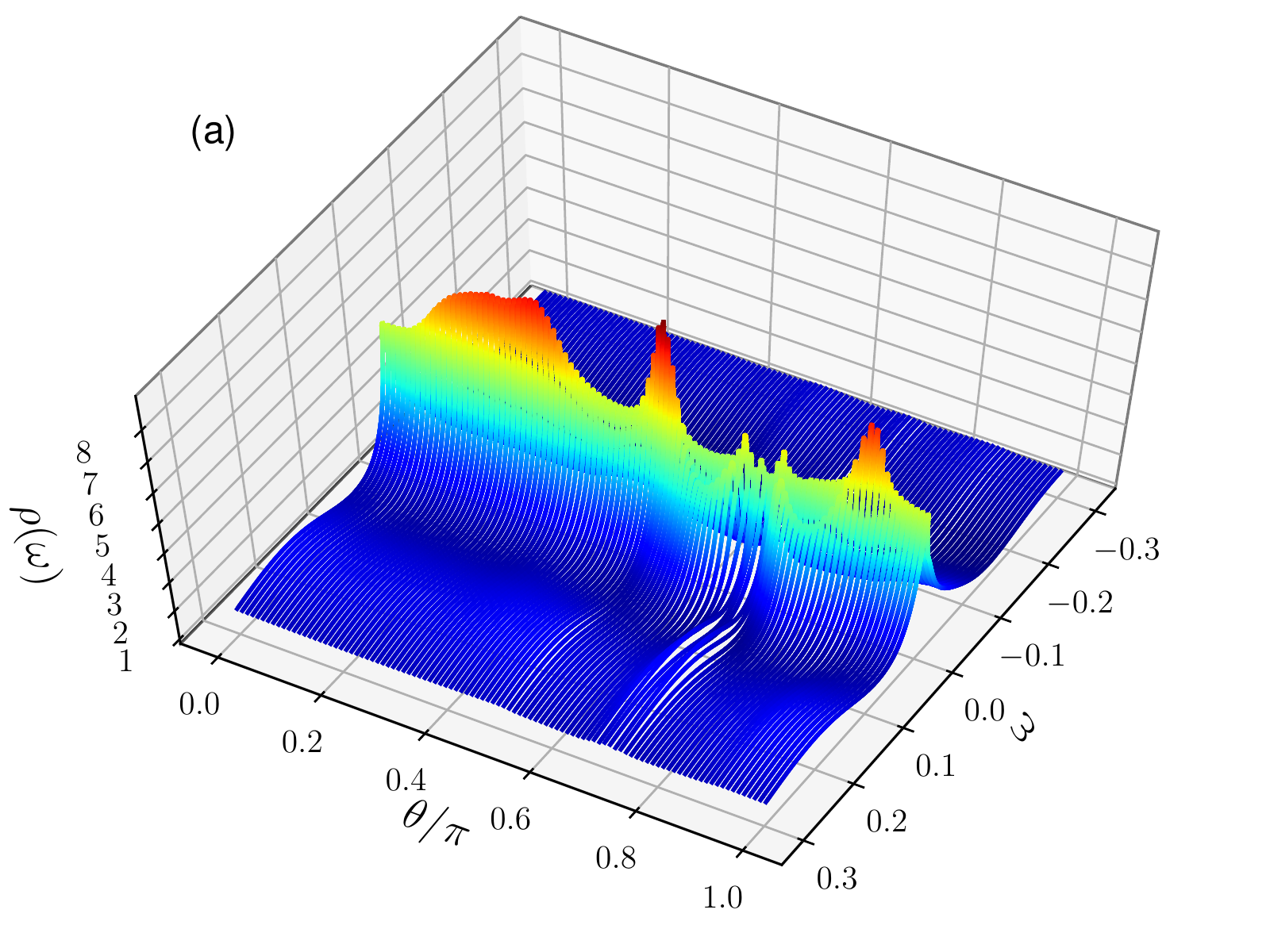}
	}
	\subfigure{\includegraphics[clip,width=1.0\columnwidth]{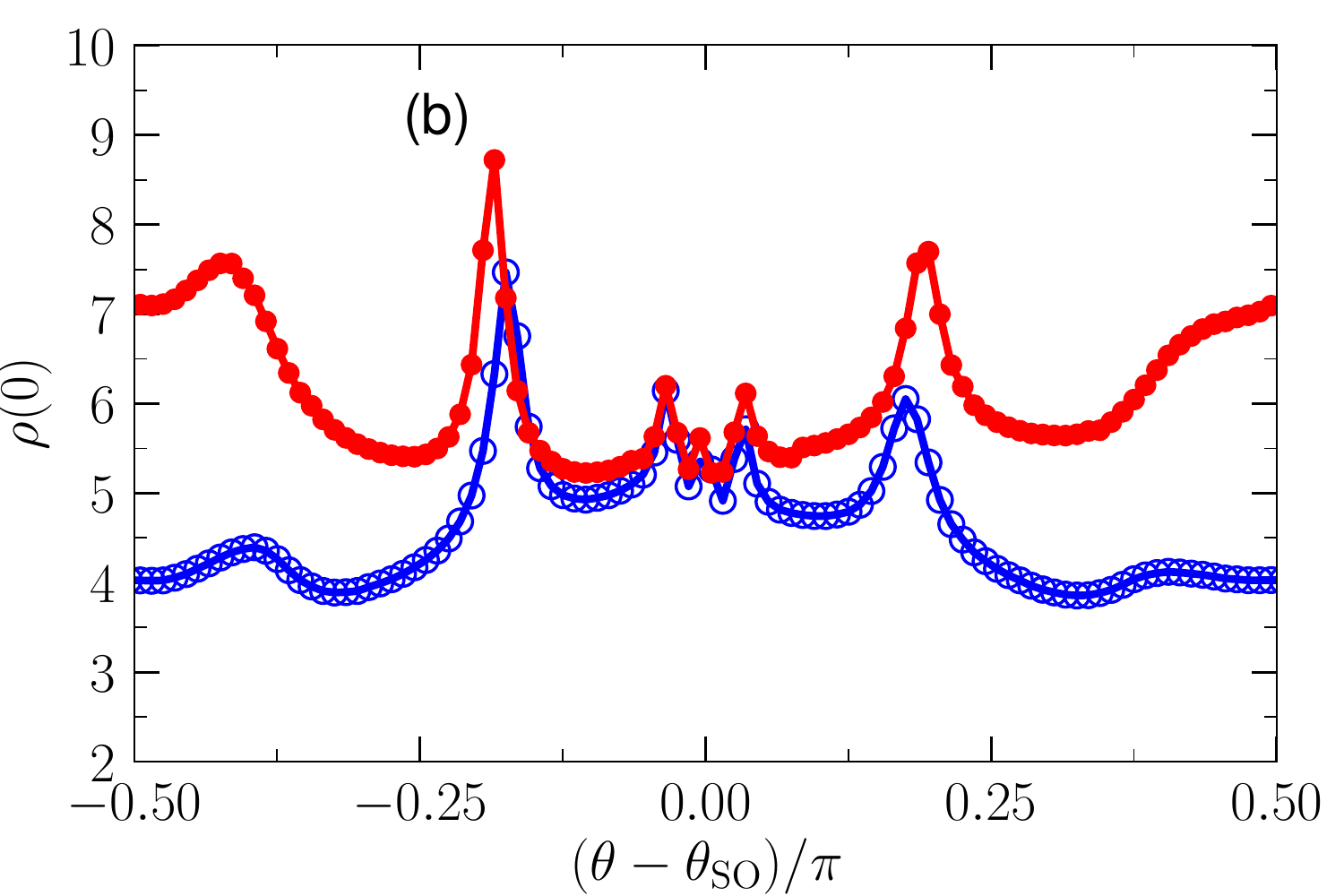}
}
\caption{\label{fig7} (a) Local density of states $\rho(\omega)$ 
	\emph{vs} $\omega$ and $\theta$, for  $\alpha=0.4$ and $\beta=0.3$ ($\theta_{\rm SO}\approx -0.295 \pi$).
	(b) $\rho(\omega=0)$ \emph{vs} $\theta - \theta_{\rm SO}$ for $g_{\rm imp }=0$ 
	(blue open symbols) and $g_{\rm imp}=0.4$ (red solid symbols), where 
	$-\nicefrac{\pi}{2} \leq \theta - \theta_{\rm SO} \leq \nicefrac{\pi}{2}$ 
	determines the orientation of the external magnetic field in the $xy$ plane 
	in relation to $\theta_{\rm SO}$. Note that the curves have a $\pi$ periodicity. 
	$g_{\rm w}=1$ for both panels.}
\end{figure}

In Fig.~\ref{figmagvsB}, 
we present how the $g_{\rm imp}=0.4$ impurity magnetization $-\langle S_i \rangle$, for $i=x,y,z$, varies 
with external field intensity ($0 \leq B \leq 0.01$), for the field applied along the $i$-axis. 
The results for $\By$ (blue curve) and $\Bx$ and $\hat{\vec{z}}$ (red curve) evolve smoothly with 
field intensity, with the $\By$ curve seemingly having plateaued around $B=0.01$. Thus, the $B=0.01$ results 
presented in the previous sections may be considered as representative, i.e., 
there is nothing special about the $B=0.01$ value. In Fig.~\ref{figKpeakvsB}, we show the spin-down projected 
Kondo peak position, denoted as $\omega_{\downarrow}^{\rm max}$, 
as a function of $\vecB$ ($0 \leq B \leq 0.01$), for $g_{\rm imp}=0.4$. As for the case of the impurity magnetization, 
Fig.~\ref{figmagvsB}, both curves evolve smoothly with external field, showing again that the $B=0.01$ value 
is representative of the physical phenomena discussed above.

\subsection{Combined effect of Rashba and Dresselhaus SOC}

We now consider the generic case with both Rashba and Dresselhaus SOC.
Based on what has been shown so far
we anticipate that an analysis of the Kondo peak height as a function
of the field direction provides information about the direction
of $\vecB_{\rm SOC}$.
Since $\theta_{\rm SOC}$ is associated with the ratio
$\nicefrac{\alpha}{\beta}$, its precise determination (e.g. using scanning
tunneling spectroscopy) in conjuction with additional measurements~\cite{Meier2007,Park2013,Knox2018} 
would give access to the absolute values of $\alpha$ and $\beta$. 
Panel (a) in Fig.~\ref{fig7} shows a 3D plot of the impurity's LDOS for
the magnetic field in the $xy$ plane as a function of the polar angle
$\theta$ between the $x$-axis and the field direction.
One can clearly see that the Kondo peak (at $\omega=0$) suffers strong variations as a function of $\theta$. 
This can be observed in more detail in Fig.~\ref{fig7}(b), which shows the impurity LDOS at the Fermi energy  
(i.e., Kondo peak height) as a function of $\theta - \theta_{\rm SO}$, the direction 
of the external magnetic field in relation to $\theta_{\rm SO}$, 
from $-\nicefrac{\pi}{2}$ to $\nicefrac{\pi}{2}$. Open (blue) symbols are for $g_{\rm imp }=0$, 
while solid (red) symbols are for $g_{\rm imp}=0.4$. 
We note that the spin symmetry of the Hamiltonian requires 
that the curves in panel (b) should be symmetric around $\theta=\theta_{\rm SO}$.  
The somewhat delicate NRG numerics at $\omega=0$ is responsible 
for the observed lack of perfect symmetry.
Two broad $\rho(0)$ maxima occur orthogonally to 
$\theta_{\rm SOC} = -\tan^{-1}\nicefrac{\alpha}{\beta}$.
This is in agreement with the results described above as 
a `revival of the Kondo peak' for $\vecB \perp \vecB_{\rm SOC}$. 
The presence of other features in the curves indicates that a 
better strategy to find $\theta_{\rm SO}$ is by exploiting 
the expected symmetry around $\theta_{\rm SO}$.
In any case, this method of finding the Rashba and Dresselhaus couplings can be used as a complementary 
technique to other proposed procedures~\cite{Meier2007,Park2013,Knox2018}.

A very interesting recent experimental result~\cite{Bommer2019} has 
shown a similar magnetic-field-revealed anisotropy in an InSb quantum wire 
proximity coupled to a superconductor. In that case, it is the superconducting gap 
that undergoes a `revival' when the magnetic field is rotated away 
from the SOC-induced effective magnetic field. 

\section{Summary and Conclusions} 
We have shown that Rashba and Dresselhaus 
SOC in a quantum wire can be investigated through their combined effect on the Kondo ground state of a quantum 
impurity coupled to the wire. Although SOC breaks the spin isotropy through the 
introduction of an effective magnetic field $\vecB_{\rm SO}$, 
this anisotropy is only manifested when 
an external magnetic field $\vecB$ is applied. In that case, the Kondo state properties, like the height of the 
Kondo peak as well as its Zeeman splitting, are strongly dependent on the relative orientation 
of $\vecB_{\rm SO}$ and $\vecB$. The maximum suppression of the Kondo 
peak occurs for $\vecB_{\rm SO} \parallel \vecB$. Since the orientation of $\vecB_{\rm SO}$ is given by 
$\theta_{\rm SO}=\tan^{-1}\nicefrac{\alpha}{\beta}$, where $\alpha$ and $\beta$ parametrize the Rashba and 
Dresselhaus interaction, determination of $\theta_{\rm SO}$ can be used to estimate 
$\nicefrac{\alpha}{\beta}$. 
Finally, it would be interesting, as a possible follow-up work, to study the role of 
the ratio $\nicefrac{g_\mathrm{imp}}{g_w}$ more systematically.

\section{Acknowledgments}
GBM acknowledges financial support from the Brazilian agency Conselho Nacional de 
Desenvolvimento Cient\'{\i}fico e Tecnol\'ogico (CNPq), processes 424711/2018-4 and 305150/2017-0. 
R.~\v{Z}. is supported by Slovenian Research Agency (ARRS) under Program P1-0044.

\bibliography{rashba}

\begin{thebibliography}{59}
\bibitem [{not({\natexlab{a}})}]{note7}%
  \BibitemOpen
\bibinfo {note} {The quantum wire
  polarization $\langle S_i^{\rm w} \rangle$ is calculated as $\sum_k \langle k
  | S_i^{\rm w} | k \rangle$, where $| k \rangle$ are the eigenstates of ${\cal
  H}_\wire$ in Eq.~\eqref{H_wire}. The sum is over the first Brillouin zone, up
  to the Fermi energy.}\BibitemShut {Stop}%
\bibitem [{not({\natexlab{c}})}]{note0}%
  \BibitemOpen
  \bibinfo {note} {More specifically, the
  spin-resolved density of states is given by $\rho_{\sigma\sigma}(\omega) =
  \Gamma _{\sigma\sigma}(\omega) /(\pi V^{2})$. Thus, $\rho_{\uparrow\uparrow}
  \propto d_0 + d_z$, $\rho_{\downarrow\downarrow} \propto d_0 - d_z$,
  resulting in $\rho=\rho_{\uparrow\uparrow} + \rho_{\downarrow\downarrow}
  \propto 2 d_0$.}\BibitemShut {Stop}%
\bibitem [{not({\natexlab{d}})}]{note9}%
  \BibitemOpen
  \bibinfo {note} {One has to be careful
  with the terminology (and the signs) when referring to the impurity spin
  polarization along direction $i$, denoted $\langle S_i \rangle$, and the
  impurity magnetization, given by $- \langle S_i \rangle$. Thus, for just a
  Zeeman term, we have that the impurity magnetization aligns with the external
  field, while the impurity spin polarization antialigns with it.}\BibitemShut
  {Stop}%
\end{thebibliography}

\end{document}